\documentclass{an}
\usepackage{times}
\usepackage{graphicx}
\usepackage{natbib}

\bibpunct{(}{)}{;}{a}{}{,}
\graphicspath{{PS/}}
\newcommand{\nc}{\newcommand}

\newcommand{\mcl}{\multicolumn}
\def\url#1{\texttt{#1}}

\def\bibcode#1{}
\nc{\cfig}[1]{\centerline{\psfig{#1}}}
\nc{\Ncol}[1]{\mcl{#1}{c}{~}}
\nc{\mcn}[1]{\mcl{#1}{l}{~}}
\nc{\mcc}[1]{\mcl{1}{c}{#1}}
\nc{\fre}{$\circ$}
\nc{\fix}{$\bullet$}
\nc{\Mpm}{$\pm$}
\nc{\eg}{e.g.\ }
\nc{\etc}{etc.}
\nc{\cf}{cf.\ }
\nc{\ie}{i.e.\ }
\nc{\eler}[2]{{\rm #1}\,{\rm \small{#2}}\,}
\nc{\asec}{\raisebox{0.5ex}{\tt "}}
\nc{\EE}[2]{$#1\times 10^{#2}$}
\nc{\todo}[1]{({\bf TODO:} #1)}
\nc{\Vg}{V$_{\rm 0}$}
\nc{\Kp}{K$_{\rm 1}$}
\nc{\Ks}{K$_{\rm 2}$}
\nc{\Tp}{T$_{\rm 1}$}
\nc{\Ts}{T$_{\rm 2}$}
\nc{\teff}{T$_{eff}$}
\nc{\pdot}{$\dot{P}$}
\nc{\AAA}{$\lambda\lambda$}
\nc{\amm}{\AA mm$^{-1}$}
\nc{\kms}{km~s$^{-1}$}
\nc{\msun}{{\rm M_\odot}}
\nc{\rsun}{{\rm R_\odot}}
\nc{\lsun}{{\rm L_\odot}}

\def\Npsr{1315}

\Volume{000}             
\Year{0000}              
\Month{00}               
\Pagespan{000}{000}      
\begin{document}

\lhead[\thepage]{A.N. Author: Title}
\rhead[Astron. Nachr./AN~{\bf XXX} (200X) X]{\thepage}
\headnote{Astron. Nachr./AN {\bf 32X} (200X) X, XXX--XXX}

\title%
{
  Distances and Other Parameters for \Npsr\ Radio Pulsar
}
\author{%
O.\,H.\ Guseinov\inst{1,2}
\and
S.\,K.\ Yerli\inst{2}
\and
S.\ Ozkan\inst{1}
\and
A.\ Sezer\inst{1}
\and
S.\,O.\ Tagiyeva\inst{3}
}

\institute{
Akdeniz Universitesi, Physics Department, 07058 Antalya, Turkey
\and
Orta Dogu Teknik Universitesi, Physics Department, 06531 Ankara, Turkey
\and
Academy of Science, Physics Institute, Baku 370143, Azerbaijan Republic
}


\date{Received {\it date will be inserted by the editor}; 
accepted {\it date will be inserted by the editor}} 

\abstract{
  In this work we have collected observational data for \Npsr\
  PSRs.
  Distances and others parameters for these PSRs were estimated.
  We present improved distance estimates for radio pulsars by
  considering importance of their physical properties and
  improvement of distribution of SFRs (star formation regions)
  in the Galaxy.
  For this purpose, both a list of accurate calibrators was
  constructed and several accurate criteria were established.
  The following values were calculated from PSRs observational
  data: luminosities at 400 Mhz and 1400 Mhz, characteristic
  times, strength of magnetic field and rate of rotation energy.
  This compilation of data is mainly necessary for statistical
  investigations and for the physical properties of neutron stars.
  The whole data is prepared in a publicly accessible web page:
  \url{http://www.xrbc.org/pulsar/}.
}
\correspondence{yerli@metu.edu.tr}

\maketitle

\section{Introduction}
\label{s:intro}

It is a well known fact that no relation has been found in pulsar
parameters to estimate their distance.
For ordinary distant stars, however, one can use the relations either
between luminosity and spectral class, or luminosity and pulsation
period to estimate their distance.
In the early days of pulsar astronomy, since origin of pulsars, mass
of their progenitor and their birth rates were not well known,
homogeneous electron density distribution was assumed.
However, later on pulsars distances have been estimated according to
the rough model of Galactic electron distribution and some natural
requirements \citep{%
	1981AJ.....86.1953M,1981AZh....58..996G,
	1992MNRAS.255..401J,1993ApJ...411..674T,
	1996TJPhy..20..275G}.
In doing this, one should also know some of the pulsar distances
independent to their dispersion measure (DM).
21 cm line of neutral Hydrogen was mainly used in choosing distance
calibrators.
However, nowadays, calibrators are chosen from members of globular
clusters (GCs) or Magellanic clouds (MC), pulsars connected to
Supernova remnants (SNRs) with well known distances and from pulsars
(PSRs), where their distances are known from other available data.

Irregularities were observed in the distribution of dust, molecular
clouds and neutral Hydrogen (HI) in the Galaxy.
It is also normal to expect irregularities in electron distribution
where the degree of irregularity is (naturally) considerably small.
Considerable variations in opacity and polarization can be observed
for stars with the same distance in a very small region of sky
($\sim$1\degr\ square) close to the Galactic plane.
This is due to a very inhomogeneous distribution of dust clouds.
For Hydrogen column density along the line of sight there are two
surveys where they studied large number of stars; one with 554 stars
\citep{1994ApJS...93..211D} and the other with 594 stars
\citep{1994ApJS...94..127F}.
They both show that irregularities in HI distribution are quite
different than the dust and molecular cloud distribution
\citep{1998A&AT...17..301A}.
The dispersion measure (DM), which is connected with the electron
distribution, changes also for pulsars of similar distances, and for
close regions of the sky.
These irregularities in electron distribution are due to contribution
of both HII regions and SNRs along the line of sight, and
gravitational potential and gas temperature distribution in the
Galaxy.
But irregularities in electron distribution is considerably smaller
than the ones in other components of interstellar medium that we have
mentioned above.
Even though these irregularities are small, there is no simple model
for Galactic electron distribution to calculate each pulsar's
distance.
Moreover, constructing a complex model which requires a lot of data
for interstellar medium and PSRs, \citep[\eg][]{1993ApJ...411..674T}
cannot avoid large errors for individual pulsars.

In order to investigate the arm structure around the Sun within a
distance of 4--5 kpc, usually objects like OB associations and open
clusters (OC) are studied.
For these objects the relative errors in estimating their distances
could reach 30\% \citep{%
	1978ApJS...38..309H,1989QB806.E37......,
	1992A&AS...94..211G,1995A&AS..109..375A}.
There is no single good method to estimate the distance of all
extended objects belonging to the arms (molecular clouds, neutral
Hydrogen clouds, SNRs and HII regions).
In determining the distances to these objects using HI 21 cm line and
Galaxy rotation models, error exceeds 30\% and it increases with
distance and in the vicinity of longitudes $l=0$\degr\ and
$l=180$\degr.
However, it is the most widely used model.
For distant X-ray sources, Hydrogen column density is used as another
method in estimating distances.
However, error in this method is also large.
Since progenitors of pulsars are massive stars, their birth places are
in the star formation regions (SFRs).
Furthermore, even though young pulsars with characteristic age of
$\tau<$\EE{5}{5} years have high space velocities, they cannot escape
from their birth places.
Thus, if number of young pulsars discovered increases and distances to
these pulsars are well known then farther away arm structures could be
studied.

Archiving radio pulsar data, dates back to 1981.
The first full catalog included 333 pulsars which covered
discoveries up to 1980 \citep{1981AJ.....86.1953M}.
The next catalog which plays an important role in pulsar astronomy
contained 706 pulsars \citep{1996unpubwork.....T}.
This one covered both old (since 1981) and new pulsars
	\citep[][and some others]{%
	1985ApJ...294L..25D,1985Natur.317..787S,
	1986ApJ...311..694S,1992MNRAS.254..177C,
	1992MNRAS.255..401J,1993ApJ...411..674T}.
This last catalog has not been updated since then.
However, individual pulsars can be reached through a publicly
accessible web page
\footnote{\small\url{http://pulsar.ucolick.org/cog/pulsars/catalog/}}.
Since 1996, several pulsar surveys have been carried out \citep{%
	1995MNRAS.274L..43J,1996MNRAS.279.1235M,
	1997ApJ...478L..95S,1998MNRAS.295..743L,
	2000MNRAS.312..698L,2001ApJ...548L.187C,
	2001ApJ...547L..37E,2001ApJ...553..801E,
	2001MNRAS.326..358E,2001ApJ...560..365E,
	2001PASA...18....1M}.
In addition to this, inner regions of SNRs have been scanned to search
for pulsars with connections to SNRs \citep{%
	1996ApJ...458..257G,1998A&A...331.1002L,1996AJ....111.2028K}.
After 1996, the following pulsars with connections to SNRs or pulsars
with confirmed association connections have been found and their
distances were accurately determined (see Table \ref{t:calib}):\\
\begin{tabular}{@{}l l@{}}
J0205+64/G130.7+3.1   &\citep{2002ApJ...568..226M},\\
J1119-6127/G292.2-0.5 &\citep{2001ApJ...554..152C},\\
                      &\citep{2001ApJ...554..161P},\\
J1124-5916/G292.0+1.8 &\citep{2002ApJ...567L..71C},\\
J1803-2137/G8.7-0.1?  &\citep{1994ApJ...434L..25F},\\
J1846-0258/G29.7-0.3  &\citep{2000ApJ...542L..37G},\\
J1952+3525/G69.0+2.7  &\citep{1990ApJ...364..178K},\\
J2229+6114/G106.6+2.9 &\citep{2001ApJ...552L.125H}.\\
\end{tabular}\\
Furthermore, Globular Clusters (GC) have been also searched for
pulsars \citep{%
	1995mpds.conf...35L,1996IAUS..174..181K,
	1996MNRAS.282..691B,
	2000ApJ...535..975C,2001ApJ...548L.171D}.
In globular cluster NGC104 (47 Tuc) 10 pulsar up to 1996 and 10 more
pulsar after 1996 have been found.
For the other known globular clusters no new pulsars were found.
However, in each globular clusters NGC 6266, NGC 6342, NGC 6397, NGC
6544 and NGC 6752 one pulsar has been found after 1996 (Table
\ref{t:calib}).
\begin{table*}
\caption{
  Pulsars for which errors in distances are not more than 30\%.
}
\label{t:calib}

\begin{tabular}{@{}l r@{}c@{}l r@{}c@{}l r@{}c@{}l r r l l@{}} \hline
Name &
\mcl{3}{c}{$l$} &
\mcl{3}{c}{b} &
\mcl{3}{c}{d} &
DM &
$n_e$ &
Location &
References \\ \hline
0024-7204W & 305 & . & 9 & -44 & . & 9 & 4 & . & 5 & 24.3 & 0.005 & GC NGC 104 (47 Tuc) & (Harr96, Hess87) \\
0045-7319 & 303 & . & 5 & -43 & . & 8 & 57 &  &  & 105.4 & 0.002 & SMC & (Feas87) \\
0205+6449 & 130 & . & 7 & 3 & . & 1 & 3 & . & 2 & 140.7 & 0.044 & SNR G130.7+3.1 & (Cami02) \\
0455-6947 & 281 & . & 2 & -35 & . & 2 & 50 &  &  & 91.0 & 0.002 & LMC & (Feas87) \\
0502-6625 & 277 & . & 3 & -35 & . & 5 & 50 &  &  & 65.0 & 0.001 & LMC & (Feas87) \\
0529-6655 & 277 & . & 2 & -32 & . & 8 & 50 &  &  & 100.0 & 0.002 & LMC & (Feas87) \\
0534+2200 & 184 & . & 6 & -5 & . & 8 & 2 &  &  & 56.8 & 0.028 & SNR G184.6-5.8 (Crab) & (Trim71) \\
0540-6919 & 279 & . & 7 & -31 & . & 5 & 50 &  &  & 146.0 & 0.003 & LMC & (Tayl96) \\
0826+2637 & 196 & . & 9 & 31 & . & 7 & 0 & . & 4 & 19.5 & 0.049 & Parallax & (Gwin86) \\
0835-4510 & 263 & . & 6 & -2 & . & 8 & 0 & . & 45 & 68.2 & 0.152 & SNR G263.9-3.3 (Vela) & (Cha99, Legg00, Guse02) \\
0922+0638 & 225 & . & 4 & 36 & . & 4 & 1 & . & 21 & 27.3 & 0.020 & Parallax & (Chat01, Foma99) \\
0953+0755 & 228 & . & 9 & 43 & . & 7 & 0 & . & 28 & 3.0 & 0.011 & Parallax & (Gwin86, Bris00) \\
1119-6127 & 292 & . & 2 & -0 & . & 54 & 7 & . & 5 & 707.4 & 0.101 & SNR G292.2-0.5 & (Craw01, Guse02, Kasp02) \\
1302-6350 & 304 & . & 2 & -0 & . & 9 & 1 & . & 3 & 146.7 & 0.113 & Sp Binary, Be & (John94) \\
1312+1810 & 332 & . & 9 & 79 & . & 8 & 18 & . & 9 & 24.0 & 0.001 & GC NGC 5024 (M53) & (Harr96, Rey98) \\
1341-6220 & 308 & . & 7 & -0 & . & 4 & 8 &  &  & 730.0 & 0.097 & SNR G308.8-0.1 & (Casw92, Guse02) \\
1456-6843 & 313 & . & 9 & -8 & . & 5 & 0 & . & 45 & 8.6 & 0.019 & Parallax & (Bail90) \\
1513-5908 & 320 & . & 3 & -1 & . & 2 & 4 & . & 2 & 253.2 & 0.060 & SNR G320.4-1.2 & (Tayl96, Guse02, Kasp02) \\
1518+0204B & 3 & . & 9 & 46 & . & 8 & 7 &  &  & 30.5 & 0.004 & GC NGC 5904 (M5) & (Harr96, Broc96b, Sand96) \\
1623-2631 & 350 & . & 9 & 15 & . & 9 & 1 & . & 8 & 62.9 & 0.035 & GC NGC 6121 (M4) & (Harr96, Cudw90) \\
1641+3627B & 59 & . & 8 & 40 & . & 9 & 7 & . & 7 & 29.5 & 0.004 & GC NGC 6205 (M13) & (Harr96, Palt98) \\
1701-30 & 353 & . & 6 & 7 & . & 3 & 5 &  &  & 114.4 & 0.023 & GC NGC 6266 (M62) & (Harr96, Dami01, Broc96a) \\
1721-1936 & 4 & . & 9 & 9 & . & 7 & 9 &  &  & 71.0 & 0.008 & GC NGC 6342 & (Harr96, Heit99) \\
1740-53 & 338 & . & 2 & -11 & . & 9 & 2 & . & 3 & 71.8 & 0.031 & GC NGC 6397 & (Harr96, Dami01, Alca87) \\
1748-2021 & 7 & . & 7 & 3 & . & 8 & 6 & . & 6 & 220.0 & 0.033 & GC NGC 6440 & (Harr96, Orto94) \\
1748-2445B & 3 & . & 85 & 1 & . & 7 & 7 &  &  & 205.0 & 0.029 & Ter 5 & (Harr96, Orto96) \\
1801-2451 & 5 & . & 2 & -0 & . & 9 & 4 & . & 5 & 289.0 & 0.064 & SNR G5.4-1.2 & (Frai94, Guse02, Kasp02) \\
1803-2137 & 8 & . & 4 & 0 & . & 1 & 3 & . & 5 & 233.9 & 0.067 & SNR G8.7-0.1 & (Frai94, Alla97, Finl94, Guse02, Kasp02) \\
1804-0735 & 20 & . & 8 & 6 & . & 8 & 7 &  &  & 186.4 & 0.027 & GC NGC 6539 & (Harr96, Arma88) \\
1807-2459 & 5 & . & 8 & -2 & . & 2 & 2 & . & 6 & 134.0 & 0.052 & GC NGC 6544 & (Harr96, Dami01, Kasp94) \\
1823-3021B & 2 & . & 8 & -7 & . & 9 & 8 &  &  & 87.0 & 0.011 & GC NGC 6624 & (Harr96, Sara94) \\
1824-2452 & 7 & . & 8 & -5 & . & 6 & 5 & . & 7 & 119.8 & 0.021 & GC NGC 6626 (M28) & (Harr96, Rees91) \\
1856+0113 & 34 & . & 6 & -0 & . & 5 & 2 & . & 8 & 96.7 & 0.035 & SNR G34.7-0.4 & (Kasp00, Guse02, Kasp02) \\
1910-59 & 336 & . & 5 & -25 & . & 6 & 4 &  &  & 34.0 & 0.009 & GC NGC 6752 & (Harr96, Dami01, Buon86) \\
1910+0004 & 35 & . & 2 & -4 & . & 2 & 6 & . & 5 & 201.5 & 0.031 & GC NGC 6760 & (Harr96, Heit99) \\
1932+1059 & 47 & . & 4 & -3 & . & 9 & 0 & . & 17 & 3.2 & 0.019 & Parallax & (Weis80, Salt79, Back82, Camp95) \\
1952+3252 & 68 & . & 8 & 2 & . & 8 & 2 &  &  & 45.0 & 0.022 & SNR G69.0+2.7 & (Tayl96, Alla97, Guse02, Kasp02) \\
2022+5154 & 87 & . & 9 & 8 & . & 4 & 1 & . & 1 & 22.6 & 0.021 & Parallax & (Camp96) \\
2129+1209H & 65 & . & 1 & -27 & . & 3 & 10 &  &  & 67.2 & 0.007 & GC NGC 7078 (M15) & (Tayl96) \\
2337+6151 & 114 & . & 3 & 0 & . & 2 & 2 & . & 8 & 58.4 & 0.021 & SNR G114.3+0.3 & (Furs93, Guse02, Kasp02) \\ \hline
\end{tabular}
\begin{tabular}{@{} l@{:~}l l@{:~}l l@{:~}l l@{:~}l @{}}
Alca87 & \citet{1987AJ.....94..917A} & Alla97 & \citet{1997TJPhy..21..688G} & Arma88 & \citet{1988AJ.....96..588A} \\
Back82 & \citet{1982ApJ...260..512B} & Bail90 & \citet{1990Natur.343..240B} & Bris00 & \citet{2000ApJ...541..959B} \\
Broc96a & \citet{1996A+A...311..778B} & Broc96b & \citet{1996AJ....111..809B} & Buon86 & \citet{1986A+AS...66...79B} \\
Cami02 & \citet{2002ApJ...571L..41C} & Camp95 & \citet{1995PhDT........20C} & Camp96 & \citet{1996ApJ...461L..95C} \\
Casw92 & \citet{1992ApJ...399L.151C} & Cha99 & \citet{1999ApJ...515L..25C} & Chat01 & \citet{2001ApJ...550..287C} \\
Craw01 & \citet{2001ApJ...554..152C} & Cudw90 & \citet{1990AJ.....99.1491C} & Dami01 & \citet{2001ApJ...548L.171D} \\
Feas87 & \citet{1987ARA+A..25..345F} & Finl94 & \citet{1994ApJ...434L..25F} & Foma99 & \citet{1999AJ....117.3025F} \\
Frai94 & \citet{1994AJ....107.1120F} & Furs93 & \citet{1993A+A...276..470F} & Guse02 & \citet{2002unpubwork.....G} \\
Gwin86 & \citet{1986AJ.....91..338G} & Harr96 & \citet{1996AJ....112.1487H} & Heit99 & \citet{1999A+A...347..455H} \\
Hess87 & \citet{1987PASP...99..739H} & John94 & \citet{1994MNRAS.268..430J} & Kasp02 & \citet{2002nssr.conf.....K} \\
Kasp94 & \citet{1994ApJ...428..713K} & Kasp00 & \citet{2000puas.conf..485K} & Legg00 & \citet{2000puas.conf..141L} \\
Orto94 & \citet{1994A+AS..108..653O} & Orto96 & \citet{1996A+A...308..733O} & Palt98 & \citet{1998MNRAS.293..434P} \\
Rees91 & \citet{1991AJ....102..152R} & Rey98 & \citet{1998AJ....116.1775R} & Salt79 & \citet{1979Natur.280..477S} \\
Sand96 & \citet{1996ApJ...470..910S} & Sara94 & \citet{1994ApJS...93..161S} & Tayl96 & \citet{1996unpubwork.....T} \\
Trim71 & \citet{1971ApJ...163L..97T} & Weis80 & \citet{1980A+A....88...84W} & \mcl{2}{c}{~} \\
\end{tabular}

\end{table*}

In early days of pulsar observations a base frequency of around 400
MHz was used in the search.
Since DM values of distant pulsars are high, 1400 MHz was used in
surveys and in search for PSRs in SNRs and GC.
As expected, the newly discovered pulsars are generally in the direction
of the Galactic center.
After 1996, no new pulsars have been found in Magellan Clouds (MC).
However, number of pulsars in GCs and number of millisecond pulsars
with known ages ($P <$ 0.1 sec and \pdot $<$ \EE{}{-16} sec/sec)
increased about 1.5 and 1.4 times, respectively.
There is an considerable increase in number of pulsars found with low
fluxes due to increase in both sensitivity of instrumentation used in
pulsar surveys and the number of detailed surveys.
For example in Arecibo's survey window (40\degr$\le l\le$65\degr;
$|b|\le 2.5$\degr) 12 new pulsars were found.
In this article our aim is to combine both old and new observational
pulsar data and to calculate their parameters.

\section{Pulsar Distances}
\label{s:distance}

Between 1970 and 1980, both the number of pulsars and the number of
pulsars connected with an object having a well known distance (\eg
Magellan clouds, some globular clusters and SNRs) were less.
In addition to this, since at that time there was insufficient
knowledge concerning Galactic electron distribution, it was difficult
to find a good distance value using the DM value of pulsars.
Thus, pulsars with distances estimated using HI line are used as an
extra distance calibrator.
It is known that it is impossible to calculate an object's distance
using HI line at 21 cm if the object's radial velocity component of
Galactic rotational velocity is small.
In addition to this, in certain directions and distances the suitable
distance to the shift of 21 cm line would be 2 instead of 1.
Uncertainty in calculating the distance with this method is not less
than 30--50\%.
Thus, in recent years, in determining calibrators for pulsars,
distance estimates calculated using the 21 cm line are not accepted as
a rule.
For this reason it was not possible to find a distance estimate
independent from a DM value for pulsars in certain directions and
distances.

In estimating pulsar distances, the model of Galactic electron
distribution by \citet{1993ApJ...411..674T} has been widely used in
recent years.
However, in estimating the pulsar distances, the approach of
\citet{1996TJPhy..20..275G} gave smaller distances than the ones
calculated using the model of \citet{1993ApJ...411..674T} for pulsars
farther than 4 kpc and with Galactic latitudes greater than about
10\degr.
To form a new model electron distribution, \citet{2001AJ....122..908G}
have published a huge pulsar list which could be used for calibrators.
We have decided to revise their distance values to use them as
calibrators.
In Table \ref{t:calib} we present 39 pulsars for which errors in
distances should not be higher than 30\%.
Since distances of pulsars from the same GC are the same, only one
pulsar from each GC has been included in the table.
Instead of presenting a long table, the pulsar table is prepared in a
publicly accessible web page (see section \ref{s:data}).
In this table, the total number of pulsars having distances
independent from the DM value is 68.
In Table \ref{t:calib} the number of pulsars is considerably smaller
than the one in the calibrator list of \citet{2001AJ....122..908G}.
In this table, one of the most important calibrators is PSRJ0835-4510
(in Vela SNR).
The distance for this pulsar has been adopted as 0.45 kpc, which was
given as 0.25 kpc by \citet{2001AJ....122..908G}.
This huge discrepancy needs some more explanation.

Recent estimates of Vela SNR are as follows.
	     d=0.25           kpc \citep{1989ApJ...342L..83O},
	     d=0.25$\pm$0.03  kpc \citep{1999ApJ...515L..25C},
	d$\sim$0.28           kpc \citep{1999A&A...342..839B} and
	     d=0.25$\pm$0.03  kpc \citep{2000Ap&SS.272..127D}.
In estimating the distance one should also consider that Vela SNR
expands in a dense environment.
Its magnetic field is
B$\sim$\EE{6}{-5} Gauss \citep{1996ApJ...460..729D} and
its explosion energy is
\EE{(1-2)}{51} erg \citep{2000Ap&SS.272..127D}.
Of course these values have really large errors, however, they are
themselves big too.
If we take into account all of these values then it is not acceptable
to have Vela at the same position with SNR G327.6+14.6 in the
$\Sigma-D$ diagram (remnant of Ia type supernova explosion at 500 pc
above the Galactic plane; \citealt{1997ApJ...481..838H}) which expands
in a dense environment of low matter density.
Thus, Vela must be close to other SNRs which expand in a dense
environment.

In the direction of the Vela remnant, none of the young open clusters
(OC) and OB associations have distances as small as 0.25 kpc \citep{%
	1989QB806.E37......,1993PAZh...19..957B,1997TJPhy..21..875A}.
The distance of OC Pismis 4 ($l$=262.7, b=-2.4) which belongs to the
nearest Vela OB2 association and is in the direction of Vela, is 0.6
kpc \citep{1995A&AS..109..375A}.
Since the progenitors of SNRs (or pulsars) are massive stars, one
would expect the Vela remnant to be closer to the star formation
region instead of a distance value of 0.25 kpc.

If the distance value of 0.45 kpc is accepted for Vela, than the
average electron density along the line of sight would be $n_e$=0.153
cm$^{-3}$.
The pulsar with the second biggest $n_e$ value (about 0.113 cm$^{-3}$)
is for PSR J1302-6350 ($l$=304.2, b=-0.9; companion is B$_{\rm e}$
type star; d=1.3 kpc; variable wind in the environment).
The next biggest $n_e$ value (0.107 cm$^{-3}$) is for PSR J1644-4569
($l$=339.2, b=-0.2).
Since luminosity of PSR J1644-4569 at 1400 MHz is bigger than any
other known pulsar we could estimate its distance as no more than 4.5
kpc.
Average value of $n_e$ for the rest of pulsars is around 0.04.
So, it is impossible to accept a value of 0.25 kpc for Vela PSR and
Vela SNR.
We could only reduce our initial distance estimate of 0.45 kpc to
0.4 kpc the most.

For PSR J1701-30 ($l$=353.6, b=7.3), \citet{2001ApJ...548L.171D} and
\citet{2001AJ....122..908G} adopted a distance value of 6.7 kpc and
they believed that the pulsar is inside the GC 6266 (M62).
If such a high distance value is adopted for the pulsar, then the
electron number density along the line of sight should considerably
be lower than the values for the pulsars in the same direction and
approximately at the same distance.
It is much more realistic to accept a distance value of 5 kpc for this
pulsar.
Space density of both HII regions in the direction of Galactic
center and SNRs, and a higher value of $n_e$ in the direction line of
sight do not allow to have a very different $n_e$ value for the
pulsars in the same direction and approximately at the same distance.
Thus a question mark is added for PSR J1701-30 while accepting it as a
calibrator due to doubts concerning in its distance value.
The distance values of pulsars in other GCs are within the error
limits of the ones given by \citep{2001AJ....122..908G}.
Distances of PSRs connected with SNRs have been studied in an another
unpublished work \citep{2002unpubwork.....G}.
Thus, their accurate distance values have been listed in Table
\ref{t:calib}.
Among the pulsars that were used as calibrators and were a member of
GC, the ones with the most varying distances were PSR J1748-2445A
and B, J1804-0735 and J1910+0004 in GCs Ter 5, NGC 6539 and NGC 6760,
respectively.
These variations are due to the fact that new distances of these GCs
are more than two times bigger than the estimates before 1996.

It is a well known fact that dynamical equilibrium could be achieved
within the old populations (both halo and disk; characteristic time is
$\approx 10^{10}$ years).
However, these populations are not in dynamical equilibrium with each
other.
Total mass of stars and gas which belongs to Galactic arms is about
1\% of the total mass of Galaxy and parameters of arm structure
changes with time.
Characteristic time of these changes is about 10$^9$ years.
On the other hand, SFRs which are far from dynamical stability have an
order of magnitude smaller ages than the characteristic time of arm
structures.
Therefore, one should not expect any coincidence between the geometric
plane of arms and the Galactic plane throughout the whole Galaxy.
SFRs might be found either below or above the Galactic plane.
Optical observations of Cepheids with high luminosities (variables
with long pulse periods) and red supergiants at a distance of
$\approx$5-10 kpc from the Sun in the direction of
$l\approx$200--330\degr\ have shown that SFRs lie about 300 pc below
the Galactic plane.
Similarly at the same distance and in the direction of
$l\approx$70--100\degr\ SFRs lie about 400 pc above the Galactic plane.
Finally, between 3 and 5 kpc distance and in the direction of
270--320\degr\ massive Cepheids and red supergiants have been located
about 150 pc below the Galactic plane (\citep{1987SvAL...13...45B}).

\begin{figure}
\resizebox{\hsize}{!}{%
\includegraphics[angle=-90]{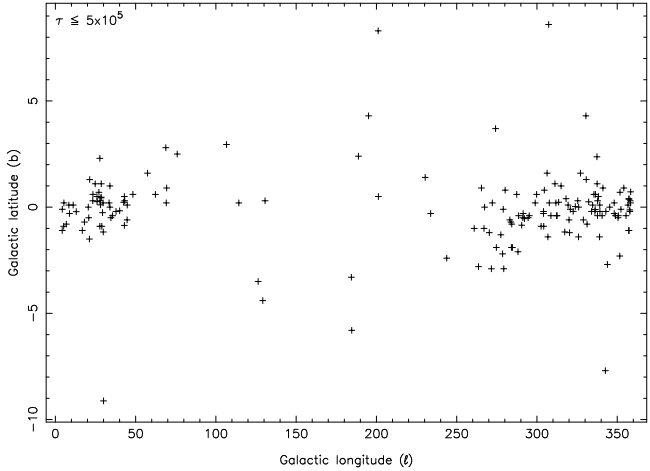}
}
\caption
{
	Galactic longitude vs latitude distribution for pulsars with
	$\tau\le$\EE{5}{5}.
}
\label{f:l-b}
\end{figure}

In Figure \ref{f:l-b}, we present $l$-b distribution of pulsars with a
characteristic time of $\tau\le$\EE{5}{5} years.
As can be seen from the Figure, in the direction of
$l\approx$260--290\degr, some young pulsar are located below the
Galactic plane.
Distance of these pulsars show that their locations coincide with the
location of Cepheids and red supergiants.
For pulsars with distances of d$>$5 kpc, average distance from the
Galactic plane is about -135 pc.
From the Figure we see a similar deviation from the Galactic plane in
the direction of $l\approx$50--80\degr.
These pulsars have an average $Z$ of about 150 pc and they probably
belong to the Perseus arm.
In distance estimation of pulsars we take into account all of these
facts (distribution of young pulsars in the direction of
0\degr$<l<$20\degr\ give rise to some inhomogeneity in pulsar surveys).

We discussed the fact that Galactic arms (SFRs) deviate from
the Galactic plane in the outer parts of the Galaxy.
However, for the inner part of the Galaxy (closer than the Sun
distance \ie about 8.5 kpc to the center) there is no evidence that
the deviation from the Galactic plane is bigger than 100 pc.
Therefore PSRs with the same age should have the same distance from
the Galactic plane because average space velocity of pulsars do not
depend on environmental conditions of a pulsar.

It is normal to neglect the influence of deviation from the Galactic
plane for pulsars older than $\approx$\EE{5}{6} years due to typical
high space velocities of pulsars (on the average between
250 \kms \citep{1997ARep...41..257A} and
450 \kms \citep{1994Natur.369..127L}).
On the other handi, the space velocity of some pulsars reaches 1000
\kms; \eg PSR J1801-2451 \citep{1991Natur.352..785F}.
But since the number of these type of pulsars are few, old pulsars
with the same age must have the same average value of $|Z|$ in all
parts of the Galaxy, except the young ones.

Radio luminosity of PSRs should not depend on their birth place and
should not considerably exceed luminosities of the strongest pulsars
(\eg Crab with a very well known distance and being the strongest
pulsar in Magellanic clouds).
Luminosity of Crab is \EE{2.6}{3} mJy kpc$^2$, 56 mJy kpc$^2$ for 400
Mhz and 1400 Mhz, respectively.
Luminosity of the strongest pulsar in Magellanic Clouds (PSR
J0529-6655) is \EE{1.4}{4} mJy kpc$^2$ at the 400 MHz (no measurement
exists for 1400 MHz).
Therefore the upper limit for luminosities of PSRs might be close to
the value of \EE{1.6}{4} mJy kpc$^2$ and \EE{3.5}{3} mJy kpc$^2$ for
400 Mhz and 1400 Mhz, respectively (spectral indices of PSRs have been
also taken into account).
In our list of \Npsr\ PSRs the strongest one is PSR J1644-4559 with
luminosities of \EE{6.29}{3} mJy kpc$^2$ and \EE{7.58}{3} mJy kpc$^2$
for 400 and 1400 MHz, respectively.

Since PSRs on the Galactic plane were born in the Galactic plane and
surveys have scanned the Galactic plane many times, most of PSRs,
especially the farthest ones, have small Galactic latitudes
($|b| < 5\degr$).
As can be seen in our calibrator table (Table \ref{t:calib}),
for 12 PSRs         $|b|>30\degr$,
for 10 PSRs $30\degr>|b|>7\degr$,
for  6 PSRs $ 7\degr>|b|>3\degr$, and only
for 11 PSRs         $|b|<3\degr$.
Thus, the calibrators in Magellan Clouds, GCs and calibrators with known
trigonometric parallax's becomes insignificant for PSRs with small $|b|$.
Only 3 from our calibrator list belong to $|b|<3$\degr\ and have
distances greater than 5 kpc.
Therefore, for the PSRs with large distance and low $|b|$ values there
are almost no calibrators.
In addition to this, for such distances it is quite difficult to judge
the electron density value.

Considering the reasons given above in adopting distances for PSRs,
the following criteria become very important:
\begin{enumerate}
\item In the direction of 40\degr$<l<$320\degr\ we see the strongest
      pulsars throughout the Galaxy.
\item For all Galactic longitudes ($l$), pulsars with equal
      characteristic times ($\tau$) must have, on the average, similar
      $|Z|$ values except PSRs with $\tau\le$\EE{5}{6} years in
      the regions where SFRs are considerably above or below the
      Galactic plane.
\item PSRs with $\tau<$\EE{5}{5} years must still be near to their
      birth places \ie in the SFRs.
\item The pulsar luminosity does not depend on $l$ and d, and it
      should not exceed the luminosity of known strongest pulsars at
      400 and 1400 MHz.
\item Electron density in the Galaxy must be correlated with the
      number density of HII regions and OB associations, and it must
      increase as one approaches to the Galactic center.
\item PSR distances must be arranged in such a way that their value
      should correspond to a suitable distance value of PSRs in Table
      \ref{t:calib} (value of DM and the direction of the PSR have to
      be taken into account).
\end{enumerate}

\section{Pulsar Data}
\label{s:data}

All of the collected parameters (both observational and calculated ones)
for \Npsr\ pulsars are given separately in a publicly accessible web
page: \url{http://www.xrbc.org/pulsar/}.
Description of each column is given in Table \ref{t:desc}.
\begin{table}
\caption{
	Description of columns in pulsar table. Catalogue Columns
	with logarithmic values are marked with `(L)' in the
	description.
}
\label{t:desc}
\begin{tabular}{@{}c @{}c l@{}}
Col.~ & Notation & Desctiption  \\ \hline
1 & Name       & Name of the pulsar \\
2 & $l$        & Galactic longitude \\
3 & b          & Galactic latitude \\
4 & Location   & Magellan clouds (MC), Globular Cluster (GC), SNR \\
 & Properties & Binary (B), Triplet (T), Glitches (G) \\
5 & DM         & Dispersion Measure \\
6 & d          & Distance \\
7 & n$_e$          & Average value of electron density along the line of sight \\
8 & F$_{1400}$ & Flux at 1400 MHz \\
9 & L$_{1400}$ & (L) Corresponding luminosity for F$_{1400}$ \\
10 & F$_{400}$  & Flux at 400 MHz \\
11 & L$_{400}$  & (L) Corresponding luminosity for F$_{400}$ \\
12 & P          & Spin Period \\
13 & \pdot      & Derivative of P \\
14 & $\dot{E}$  & (L) Rate of rotation energy loss \\
15 & B          & (L) Magnetic field \\
16 & $\tau$     & (L) Characteristic time \\
\end{tabular}
\end{table}

\section{Acknowledgements}
\label{s:ack}

This work was supported in part by T\"UB\.ITAK (Turkish Scientific and
Research Council) under grant TBAG--\c{C}G4.
This research has made use of NASA's Astrophysics Data System
Bibliographic Services.

\bibliographystyle{plainnat}
\bibliography{abbrev,pulsar}

\end{document}